\documentclass[preprint,aps]{revtex4}
\usepackage{graphicx}
\usepackage{dcolumn}
\usepackage{bm}
\usepackage{epsf}

\newcommand{\nll}{\ll \hspace{-3pt}}
\newcommand{\ngg}{\hspace{-3pt} \gg}
\newcommand{\nln}{< \hspace{-3pt}}
\newcommand{\ngn}{\hspace{-3pt} >}

\begin{document}

\title{NON-MORKAVIAN QUANTUM STOCHASTIC EQUATION FOR TWO COUPLED OSCILLATORS}

\author{
\firstname{E.X.}~\surname{Alpomishev}$^{a,*}$,
\firstname{Z.}~\surname{Kanokov}$^{a, b, **}$ }

\affiliation{ $^{a)}$
{\rm Institute of Nuclear Physics, Tashkent, Uzbekistan} \\
$^{b)}$ {\rm National University of Uzbekistan, Tashkent, Uzbekistan}\\
$^{*}$ {\rm E-mail: }{\bf erkin.alpomishev@yandex.com}\\
$^{**}$  {\rm E-mail: }{\bf zokirjon@yandex.com} }

\begin{abstract}
{\scriptsize The system of nonlinear Langevin equations was obtained
by using Hamiltonian's operator of two coupling quantum oscillators
which are interacting with heat bath. By using the analytical
solution of these equations, the analytical expressions for
transport coefficients was found. Generalized Langevin equations and
fluctuation-dissipation relations are derived for the case of a
nonlinear non-Markovian noise. The explicit expressions for the
time-dependent friction and diffusion coefficients are presented for
the case of linear couplings in the coordinate between the
collective two coupled harmonic oscillators  and heat bath.}

\end{abstract}


\maketitle

\newpage

\section{INTRODUCTION}

Nowadays, one of the intensively developing topics of theoretical
and mathematical physics is the non-equilibrium quantum theory.  The
study of the dynamics of open systems is directed towards derivation
of transport equations and finding transport coefficients which they
include. Many works are devoted to developing of formalism for the
description of statistical and dynamical behavior of open systems.
Powerful apparatus for solving complicated statistical problems of
open systems is the theory of Markovian random processes and
diffusion type processes, which has the origin of Brownian motion.
However, the use of  models of Markovian random process in many
cases is quite rough, and in some cases - actually inapplicable. In
that reason, designing of mathematical methods to consider
non-Markovian random processes becomes natural and realistic. One of
the possibilities of mathematical representation of non-Markovian
process is using of integro-differential equations rather than
differential equations. This kind of approach in essence allows to
take into account the memory of the system when random process
exists in it. An important problem in the theory of quantum open
systems is the study of reduced (i.e. averaged over the reservoir
state) dynamics of the system. In this case we usually suppose that
reservoir is in equilibrium state. Reduced dynamics is described by
the master equations for reduced system density matrix or for time
evolution of system's observables averaged over the reservoir's
state. Exact master equations include effects of memory and are
complicated for practical study. Study of behaviors of dissipative
quantum non-Markovian system beyond weak coupling or high
temperatures draws an interest into exact solvable models
\cite{Difc,Kalan,MyAllDM,MyAllQDA,MyAllNT,obzor,Leggett,Katia,Weiss}.
In these models the internal subsystem (i.e., reservoir) is
represented by a set of harmonic oscillators, whose interaction with
a collective subsystem of harmonic oscillators is realized by the
linear coupling between coordinates. Density of oscillators and
coupling constants between internal and collective subsystems are
chosen so that equations of motion for averages to be consistent
with the classical formalism. Among quantum transport equations one
can recommend the phenomenological Lindblad equation
\cite{Lindblad}. This is a deterministic equation, which can be
obtained by averaging of stochastic Langevin equation by the
controlling quantum noise. In kinetic theory, Langevin's method
significantly simplifies the calculation of non-equilibrium quantum
and thermal fluctuations and provides a clear description of both
Markovian and non-Markovian dynamics of the process. The description
below is devoted to the elegant method to obtain non-stationary
friction and diffusion coefficients for subsystem in case of
arbitrary damping temperature. The transport coefficients also
includes non-Markovian effects. As a starting point, Langevin
approach is used, which is widely used for considering fluctuation
and dissipation effects in macroscopic systems.

 A quantum oscillator coupled to a heat bath is a very important and useful problem for
many processes dealing with dynamics of open quantum systems
\cite{Dorofeyev}. In this work the problem of two coupled quantum
oscillators interacting with ensembles of harmonic oscillators is
considered.

\section{GENERALIZED NON-MARKOVIAN QUANTUM LANGEVIN  EQUATIONS}

Let us define the microscopic Hamiltonian $H$ of the total system
(internal subsystem plus collective subsystem), which will be used
to obtain non-Markovian quantum stochastic Langevin equations and
time-dependent transport coefficients for the collective subsystem.
In  a quantum Hamiltonian was constructed for the systems, which is explicitly
dependent on the collective coordinates $R$ and $\beta$,  canonically conjugate collective momentums $P$ and $P_{\beta}$  and
internal degrees of freedom
\begin{eqnarray}
H&=&H_s+H_b+H_{sb}
\label{1}
\end{eqnarray}
\begin{eqnarray}
H_s&=&\frac{P^2}{2m_1}+\eta \frac{m_1\omega_1^2 R^2}{2}+\frac{P_\beta^2}{2m_2}+\frac{m_2\omega_2^2 \beta^2}{2} + g_{R\beta}\left(R\cdot\beta\right)\nonumber\\
H_b&=&\sum_{\nu}^{}\hbar\omega_\nu b_\nu^{\dagger}b_\nu \\
H_{sb}&=&\sum_{\nu}^{}\left(\alpha_{\nu}R+ g_{\nu}\beta
\right)\left(b_{\nu}^{\dagger}+b_{\nu} \right)\nonumber \label{1_1a}
\end{eqnarray}
where $\eta=\pm1$. The coupling parameters $\alpha_{\nu}$ and
$g_{\nu}$ are
\begin{eqnarray}
\alpha_{\nu}^{2}=\frac{\lambda_{1}\Gamma_{\nu}^{2}}{\hbar}, \, \, \,
g_{\nu}^{2}=\frac{\lambda_{2}\Gamma_{\nu}^{2}}{\hbar}
\end{eqnarray}
where $\lambda_1$ and $\lambda_{2}$ are parameters which measure the
average strengths of the interactions and $\Gamma_{\nu}$ are the
coupling constants. $b_\nu^+$ and $b_\nu$ are the phonon production
and annihilation operators that describe internal excitations of the
system with energy $\hbar \omega_\nu$. For the sake of simplicity,
we omit the signs of the operators. The quantities $H_{s}$ and
$H_{b}$ are the Hamiltonians of the collective and the internal
subsystem respectively. The quantity $H_{sb}$ describes coupling of
the collective motion with the internal excitations and is a source
of dissipative terms appearing in the equations for the operators of
the collective variables.

Using Hamiltonian (\ref{1}), we obtain a system of quantum
Heisenberg equations for the operators related to the collective and
internal motion
\begin{eqnarray}\nonumber
\dot R(t)&=&\frac{i}{\hbar}[H,R]=\frac{P(t)}{m_1}\nonumber\\
\dot{\beta}(t)&=&\frac{i}{\hbar}[H,\beta]= \frac{P_{\beta}(t)}{m_2}\nonumber\\
\dot P(t)&=&\frac{i}{\hbar}[H,P]=-\eta m_1 \omega_1^2
R(t)-\sum_{\nu}^{}\alpha_{\nu}\left(b_{\nu}^{\dagger}(t) +
b_{\nu}(t)\right)-g_{R\beta}\beta(t)\nonumber\\
\dot
P_{\beta}(t)&=&\frac{i}{\hbar}[H,P_{\beta}]=-m_2\omega_2^2\beta(t)-
\sum_{\nu}^{}g_{\nu}\left(b_{\nu}^{\dagger}(t) +
b_{\nu}(t)\right)-g_{R\beta}R(t) \label{1_2}
\end{eqnarray}
and
\begin{eqnarray}
\dot b_\nu^{\dagger}(t)=\frac{i}{\hbar}[H,b_\nu^{\dagger}]= i\omega_\nu b_\nu^{\dagger}(t)+\frac{i}{\hbar}\left[\alpha_{\nu}R(t)+g_{\nu}\beta(t)\right],\nonumber\\
\dot b_\nu(t)=\frac{i}{\hbar}[H,b_\nu]= -i\omega_\nu
b_\nu(t)-\frac{i}{\hbar}\left[\alpha_{\nu}R(t)+g_{\nu}\beta(t)\right]
\label{1_3}
\end{eqnarray}
The solutions of Eqs.(\ref{1_3}) are
\begin{eqnarray}
b_\nu^{\dagger}(t)=f_\nu^{\dagger}(t)-\frac{1}{\hbar\omega_{\nu}} \left(\alpha_{\nu}R(t)+g_{\nu}\beta(t) \right)+\frac{1}{\omega_{\nu}}\int\limits_{0}^{t}d\tau e^{i\omega_{\nu}\left(t-\tau\right)} \left[\alpha_{\nu}\dot R(t)+ g_{\nu} \dot \beta(t) \right],\nonumber\\
b_\nu(t)=f_\nu(t)-\frac{1}{\hbar\omega_{\nu}}
\left(\alpha_{\nu}R(t)+g_{\nu}\beta(t)
\right)+\frac{1}{\omega_{\nu}}\int\limits_{0}^{t}d\tau
e^{-i\omega_{\nu}\left(t-\tau\right)} \left[\alpha_{\nu}\dot R(t)+
g_{\nu} \dot \beta(t) \right] \label{1_4}
\end{eqnarray}
Therefore,
\begin{eqnarray}
b_{\nu}^{\dagger}(t)+b_{\nu}(t)&=f_{\nu}^{\dagger}(t)+f_{\nu}(t)-\frac{2}{\hbar \omega_{\nu}}\left(\alpha_{\nu}R(t)+g_{\nu}\beta(t)\right)\\
&+\frac{2}{\hbar\omega_{\nu}}\int\limits_{0}^{t}d\tau\left[\alpha_{\nu}\dot{R}(t)+g_{\nu}\dot{\beta}(t)\right]\cos\left(\omega_{\nu}(t-\tau)\right)
\label{bnu}
\end{eqnarray}
where
\begin{eqnarray}
f_{\nu}^{\dagger}(t)&=&\left[b_{\nu}^{\dagger}(0)+\frac{\alpha_{\nu}R(0)}{\hbar\omega_{\nu}}+
\frac{g_{\nu}\beta(0)}{\hbar\omega_{\nu}}\right]e^{i\omega_{\nu}t}\nonumber\\
f_{\nu}(t)&=&\left[b_{\nu}(0)+\frac{\alpha_{\nu}R(0)}{\hbar\omega_{\nu}}+\frac{g_{\nu}\beta(0)}{\hbar\omega_{\nu}}\right]e^{-i\omega_{\nu}t}
\end{eqnarray}
Substituting Eq.(\ref{bnu}) into Eqs.(\ref{1_2}), we eliminate the bath variables from the equations of motion
of the collective subsystem and obtain the nonlinear integro-differential stochastic dissipative equations
\begin{eqnarray}\nonumber
\dot{R}(t)&=&\frac{P(t)}{m_1} \nonumber\\
\dot{\beta}(t)&=&\frac{P_{\beta}(t)}{m_2} \nonumber\\
\dot{P}(t)&=&-\eta
m_1\omega_1^{2}R(t)-g_{R\beta}\beta(t)-F_1(t)+2\sum_{\nu}^{}\frac{1}{\hbar\omega_{\nu}}\left(
\alpha_{\nu}^{2}R(t)+
\alpha_{\nu}g_{\nu}\beta(t)\right)-\nonumber\\
&-&2\sum_{\nu}^{}\frac{1}{\hbar\omega_{\nu}}\int\limits_{0}^{t}d\tau\left[\alpha_{\nu}\dot{R}(\tau)+
\alpha_{\nu}g_{\nu}\dot{\beta}(\tau)\right]\cos\left(\omega_{\nu}(t-\tau)\right)\nonumber\\
\dot{P}_{\beta}(t)&=&-m_2\omega_2^{2}\beta(t)-g_{R\beta}R(t)-F_2(t)+2\sum_{\nu}^{}\frac{1}{\hbar\omega_{\nu}}\left(
\alpha_{\nu}g_{\nu}R(t)+
g_{\nu}^2 \beta(t)\right)-\nonumber\\
&-&2\sum_{\nu}^{}\frac{1}{\hbar\omega_{\nu}}\int\limits_{0}^{t}d\tau\left[\alpha_{\nu}g_{\nu}\dot{R}(\tau)+
g_{\nu}\dot{\beta}(\tau)\right]\cos\left(\omega_{\nu}(t-\tau)\right)
\label{teng}
\end{eqnarray}
The presence of the integral parts in these equations indicates
the non-Markovian character of the system. Since in com-parison with Refs.\cite{Leggett,Weiss} we do not introduce the counter-term in the
Hamiltonian, the stiffnesses of the potentials are
renormalized in the equations above. Due to the operators
\begin{eqnarray}
F_{1}(t)&=&\sum_{\nu}^{}F_{\alpha}^{\nu}(t)=\sum_{\nu}^{}\alpha_{\nu}\left(f_{\nu}^{\dagger}(t)+f_{\nu}(t)\right)\nonumber\\
F_{2}(t)&=&\sum_{\nu}^{}F_{g}^{\nu}(t)=\sum_{\nu}^{}g_{\nu}\left(f_{\nu}^{\dagger}(t)+f_{\nu}(t)\right)\nonumber
\end{eqnarray}
which play the role of random forces in the coordinates,
Eqs.(\ref{teng}) can be called the generalized nonlinear quantum
Langevin equations. Following the usual procedure of statistical
mechanics, we identify these operators as fluctuations because of
the uncertainty in the initial conditions for the bath operators.
\begin{eqnarray}\nonumber
\dot{R}(t)&=&\frac{P(t)}{m_1} \nonumber\\
\dot{\beta}(t)&=&\frac{P_{\beta}(t)}{m_2} \nonumber\\
\dot{P}(t)&=&-\left(\eta m_1\omega_1^{2}-\bigtriangleup_{1}\right)R(t)-\left(g_{R\beta}-\bigtriangleup_{2}\right)\beta(t)-F_1(t)-\frac{1}{m_1}\int\limits_{0}^{t}d\tau K_1(t-\tau)P(\tau)-\nonumber\\
&-&\frac{1}{m_2}\int\limits_{0}^{t}d\tau K_2(t-\tau)P_{\beta}(\tau)
\nonumber\\
\dot{P}_{\beta}(t)&=&-\left( g_{R\beta}-\bigtriangleup_{3}\right)R(t)-\left(m_2\omega_2^{2}-\bigtriangleup_{4}\right)\beta(t)-F_2(t)-\frac{1}{m_1}\int\limits_{0}^{t}d\tau K_3(t-\tau)P(\tau)-\nonumber\\
&-&\frac{1}{m_2}\int\limits_{0}^{t}d\tau K_4(t-\tau)P_{\beta}(\tau)
\label{1_5}
\end{eqnarray}
where
\begin{eqnarray}\nonumber
\bigtriangleup_{1}=\sum_{\nu}^{}\frac{2\alpha_{\nu}^2}{\hbar\omega_{\nu}},
\, \, \,
\bigtriangleup_{2}=\bigtriangleup_{3}=\sum_{\nu}^{}\frac{2\alpha_{\nu}g_{\nu}}{\hbar\omega_{\nu}},
\, \, \,
\bigtriangleup_{4}=\sum_{\nu}^{}\frac{2g_{\nu}^{2}}{\hbar\omega_{\nu}}\nonumber
\end{eqnarray}
\begin{eqnarray}\nonumber
K_{1}(t-\tau)&=&\sum_{\nu}^{}\frac{2 \alpha_{\nu}^2}{\hbar\omega_{\nu}}\cos\left(\omega_{\nu}(t-\tau)\right)\nonumber\\
K_{2}(t-\tau)&=&K_{3}(t-\tau)=\sum_{\nu}^{}\frac{2\alpha_{\nu}g_{\nu}}{\hbar\omega_{\nu}}\cos\left(\omega_{\nu}(t-\tau)\right)\nonumber\\
K_{4}(t-\tau)&=&\sum_{\nu}^{}\frac{2g_{\nu}^2}{\hbar\omega_{\nu}}\cos\left(\omega_{\nu}(t-\tau)\right)\nonumber
\end{eqnarray}

In equations of motion (\ref{1_5}), the dissipative kernels
$K_{1}(t-\tau)$, $K_{2}(t-\tau)$, $K_{3}(t-\tau)$ and
$K_{4}(t-\tau)$ are separated in the terms proportional to $\dot R$,
$\dot{\beta}$ and $\dot{P}$, $\dot P_{\beta}$
\cite{Katia,MyYAF1,MyTMF1}. These kernels depend on the coefficients
of $H_{sb}$. Since the dissipative kernels do not depend on the
number of phonons, they do not depend on the bath temperature $T$
either. The temperature and the fluctuation enter into the
consideration of the dynamics of $R$, $\beta$ and $P$, $P_{\beta}$
via the distributions of the initial conditions for the internal
system. The explicit expressions for the dissipative kernels
$K_{1}(t-\tau)$, $K_{2}(t-\tau)$, $K_{3}(t-\tau)$ and
$K_{4}(t-\tau)$ and for the operators
$F_1(t)=\sum\limits_{\nu}^{}F_{\alpha}^\nu (t) $ and
$F_2(t)=\sum\limits_{\nu}^{}F_{g}^\nu (t) $ in (\ref{1_5}), which
play the role of the random $P$ and $P_{\beta}$, forces, were
obtained in \cite{Difc}.

In statistical physics the operators  $F_{\alpha}^\nu (t) $ and
$F_{g}^\nu(t) $ are identified as usual with the fluctuations due to
the uncertainty of the initial conditions for the bath operators. To
determine statistical properties of these fluctuations, we consider
an ensemble of initial conditions in which $R(0)$, $\beta(0)$,
$P(0)$ and  $P_{\beta}(0)$ are given, and the initial bath operators
are chosen from the canonical ensemble \cite{Katia,MyYAF1,MyTMF1}.
In this ensemble the fluctuations $F_{\alpha}^\nu (t)$ and
$F_{g}^\nu (t)$ are Gaussian distributions and have zero mean values
\begin{eqnarray}
\nll F^\nu_{\alpha}(t)\ngg =\nll F^\nu_{g}(t)\ngg =0 \label{1_8}
\end{eqnarray}
and nonzero second moments. The symbol $\nll...\ngg$ denotes the
average over the bath variables. The Gaussian distribution of the
random forces corresponds to the case where the bath is a set of
harmonic oscillators \cite{Kampen,Gardiner,Weiss}. To calculate the
correlation functions of the fluctuations, we will use the bath with
the Bose-Einstein statistics
\begin{eqnarray}
\nll f_\nu^+(t)f_{\nu'}^+(t')\ngg &=&\nll f_\nu(t)f_{\nu'}(t')\ngg =0,\nonumber\\
\nll f_\nu^+(t)f_{\nu'}(t') \ngg &=&
\delta_{\nu,\nu'}n_\nu e^{i\omega_\nu(t-t')},\nonumber\\
\nll f_\nu(t)f_{\nu'}^+(t') \ngg &=&\delta_{\nu,\nu'}(n_\nu+1)e^{-i\omega_\nu(t-t')}, \label{1_9}
\end{eqnarray}
where $n_\nu=[\exp (\hbar\omega_\nu /T) - 1]^{-1}$ are the
temperature occupation numbers for phonons.

Thus, a system of generalized nonlinear Langevin equations
(\ref{1_5}) is obtained. The presence of the integral terms in the
equations of motion means that the non-Markovian system remembers
the motion over the trajectory prior to the time $t$. Analytical
solution is possible if the functionals in (\ref{1_5}) are replaced
by their mean values considered to be weakly varying in time $t$ and
the renormalized potential is approximated by the harmonic (or
inverted) oscillator. In this case, we have a system of generalized
Langevin equations with dissipative memory kernels. We will solve
them using the Laplace transform ${\cal L}$  to obtain linear
equations for images.
\begin{eqnarray}
&&s R(s)-\frac{P(s)}{m_1}=R(0)\nonumber\\
&&s\beta(s)-\frac{P_{\beta}(s)}{m_2}=\beta(0)\nonumber\\
&&\left(\eta m_1\omega_{1}^2-\bigtriangleup_{1}\right)R(s)+\left(g_{R\beta}-\bigtriangleup_{2}\right)\beta(s)+
\left(s+\frac{1}{m_1}K_{1}(s)\right)P(s)\nonumber\\
&&+\frac{1}{m_2}K_2(s)P_{\beta}(s)=P(0)-F_1(s)\nonumber\\
&&\left(g_{R\beta}-\bigtriangleup_{3}\right)R(s)+\left(m_2\omega_{2}^2-\bigtriangleup_{4}\right)\beta(s)+
\frac{1}{m_1}K_{3}(s)P(s)\nonumber\\
&&+\left(s+\frac{1}{m_2}K_4(s)\right)P_{\beta}(s)=P_{\beta}(0)-F_2(s)
\label{laplas}
\end{eqnarray}
The above originals
can be found using the residue theorem, and the solutions $R(t)$, $\beta_(t)$, $P(t)$ and $P_{\beta}(t)$ can be written down in terms of the
roots $s_i$ of the equation
\begin{eqnarray}
d(s)&=&s^2\left(s+\frac{1}{m_1}K_{1}(s)\right)\left(s+\frac{1}{m_2}K_{4}(s)\right)-\frac{1}{m_1m_2}s^2K_{2}(s)K_{3}(s)+\nonumber\\
&+&\frac{s}{m_1}\left[\left(\eta m_1\omega_{1}^2-\bigtriangleup_{1}\right)\left(s+\frac{1}{m_2}K_{4}(s)\right)-\left(g_{R\beta}-\bigtriangleup_{3}\right)\frac{1}{m_2}K_{2}(s)\right]+\nonumber\\
&+&\frac{s}{m_2}\left[\left(m_2\omega_{2}^2-\bigtriangleup_{4}\right)\left(s+\frac{1}{m_1}K_{1}(s)\right)-\left(g_{R\beta}-
\bigtriangleup_{2}\right)\frac{1}{m_1}K_{3}(s)\right]+\nonumber\\
&+&\frac{1}{m_1m_2}\left[\left(\eta
m_1\omega_{1}^2-\bigtriangleup_{1}\right)\left(m_2\omega_{2}^2-
\bigtriangleup_{4}\right)-\left(g_{R\beta}-\bigtriangleup_{2}\right)\left(g_{R\beta}-\bigtriangleup_{3}\right)\right]=0
\label{d_s}
\end{eqnarray}
Expressions for the images yield explicit expressions for the
originals
\begin{eqnarray}
R(t)&=A_{1}(t)R(0)+A_{2}(t)\beta(0)+A_{3}(t)P(0)+A_{4}(t)P_{\beta}(0)-I_{1}(t)-I_{1}^{'}(t)\nonumber\\
\beta(t)&=B_{1}(t)R(0)+B_{2}(t)\beta(0)+B_{3}(t)P(0)+B_{4}(t)P_{\beta}(0)-I_{2}(t)-I_{2}^{'}(t)\nonumber\\
P(t)&=C_{1}(t)R(0)+C_{2}(t)\beta(0)+C_{3}(t)P(0)+C_{4}(t)P_{\beta}(0)-I_{3}(t)-I_{3}^{'}(t)\nonumber\\
P_{\beta}(t)&=D_{1}(t)R(0)+D_{2}(t)\beta(0)+D_{3}(t)P(0)+D_{4}(t)P_{\beta}(0)-I_{4}(t)-I_{4}^{'}(t)
\label{inv_lap}
\end{eqnarray}
where
\begin{eqnarray}
I_{R}(t)=\int\limits_{0}^{t}A_{3}(\tau)F_{1}(t-\tau)d\tau;
I_{R}^{'}(t)=\int\limits_{0}^{t}A_{4}(\tau)F_{2}(t-\tau)d\tau;\nonumber\\
I_{\beta}(t)=\int\limits_{0}^{t}B_{3}(\tau)F_{1}(t-\tau)d\tau;
I_{\beta}^{'}(t)=\int\limits_{0}^{t}B_{4}(\tau)F_{2}(t-\tau)d\tau;\nonumber\\
I_{P}(t)=\int\limits_{0}^{t}C_{3}(\tau)F_{1}(t-\tau)d\tau;
I_{P}^{'}(t)=\int\limits_{0}^{t}C_{4}(\tau)F_{2}(t-\tau)d\tau;\nonumber\\
I_{P_{\beta}}(t)=\int\limits_{0}^{t}D_{3}(\tau)F_{1}(t-\tau)d\tau;
I_{P_{\beta}}^{'}(t)=\int\limits_{0}^{t}D_{4}(\tau)F_{2}(t-\tau)d\tau;
\end{eqnarray}
where the coefficients are defined as
\begin{eqnarray*}
A_{1}(t)&=&{\cal L}^{-1}\left[\frac{1}{d(s)}\left(s\left(s+\frac{1}{m_1}K_{1}(s)\right)\left(s+\frac{1}{m_2}K_{4}(s)\right)+\frac{1}{m_2}\left(m_2\omega_{2}^{2}-
\bigtriangleup_{4}\right)\left(s+\frac{1}{m_1}K_{1}(s)\right)-\right.\right.\nonumber\\
&&-\left.\left.\frac{1}{m_1m_2}K_{3}(s)\left(g_{R\beta}-\bigtriangleup_{2}+sK_{2}(s)\right)\right)\right]\\
A_{2}(t)&=&{\cal
L}^{-1}\left[\frac{1}{d(s)}\left(\frac{1}{m_1m_2}\left(m_2\omega_{2}^{2}-\bigtriangleup_{4}\right)K_{2}(s)-\frac{1}{m_1}\left(g_{R\beta}-\bigtriangleup_{2}\right)
\left(s+\frac{1}{m_{2}}K_{4}(s)\right)\right)\right]\\
A_{3}(t)&=&{\cal L}^{-1}\left[\frac{1}{d(s)}\left(\frac{1}{m_1m_2}\left(m_2\omega_{2}^{2}-\bigtriangleup_{4}\right)+\frac{1}{m_1}s\left(s+\frac{1}{m_{2}}K_{4}(s)\right)\right)\right]\\
A_{4}(t)&=&-{\cal L}^{-1}\left[\frac{1}{d(s)}\left(\frac{1}{m_1m_2}\left(g_{R\beta}-\bigtriangleup_{2}\right)+\frac{s}{m_1m_2}K_{2}(s)\right)\right]\\
B_{1}(t)&=&{\cal
L}^{-1}\left[\frac{1}{d(s)}\left(\frac{1}{m_1m_2}\left(\eta
m_1\omega_{1}^{2}-\bigtriangleup_{1}\right)K_{3}(s)-\frac{1}{m_2}\left(g_{R\beta}-\bigtriangleup_{3}\right)
\left(s+\frac{1}{m_{1}}K_{1}(s)\right)\right)\right]\\
B_{2}(t)&=&{\cal
L}^{-1}\left[\frac{1}{d(s)}\left(s\left(s+\frac{1}{m_1}K_{1}(s)\right)\left(s+\frac{1}{m_2}K_{4}(s)\right)+\frac{1}{m_1}\left(\eta
m_1\omega_{1}^{2}-
\bigtriangleup_{1}\right)\left(s+\frac{1}{m_2}K_{4}(s)\right)-\right.\right.\\
&&\left.\left.-\frac{1}{m_1m_2}K_{2}(s)\left(g_{R\beta}-\bigtriangleup_{3}+sK_{3}(s)\right)\right)\right]\\
B_{3}(t)&=&-{\cal L}^{-1}\left[\frac{1}{d(s)}\left(\frac{1}{m_1m_2}\left(g_{R\beta}-\bigtriangleup_{3}\right)+\frac{s}{m_1m_2}K_{3}(s)\right)\right]\\
B_{4}(t)&=&{\cal L}^{-1}\left[\frac{1}{d(s)}\left(\frac{1}{m_1m_2}\left(\eta m_1\omega_{1}^{2}-\bigtriangleup_{1}\right)+\frac{1}{m_2}s\left(s+\frac{1}{m_{1}}K_{1}(s)\right)\right)\right]\\
C_{1}(t)&=&{\cal L}^{-1}\left[\frac{1}{d(s)}\left(\frac{s}{m_{2}}\left(g_{R\beta}-\bigtriangleup_{3}\right)K_{2}(s)-s\left(\eta m_{1}\omega_{1}^{2}-\bigtriangleup_{1}\right)\left(s+\frac{1}{m_2}K_{4}(s)\right)+\right.\right.\\
&&\left.\left.+\frac{1}{m_2}\left(g_{R\beta}-\bigtriangleup_{2}\right)\left(g_{R\beta}-\bigtriangleup_{3}\right)-\frac{1}{m_{2}}\left(\eta
m_1\omega_{1}^{2}-
\bigtriangleup_{1}\right)\left(m_{2}\omega_{2}^{2}-\bigtriangleup_{4}\right)\right)\right]\\
C_{2}(t)&=&{\cal
L}^{-1}\left[\frac{1}{d(s)}\left(\frac{s}{m_2}\left(m_2\omega_{2}^{2}-\bigtriangleup_{4}\right)K_{2}(s)-s\left(g_{R\beta}-
\bigtriangleup_{2}\right)\left(s+\frac{1}{m_{2}}K_{4}(s)\right)\right)\right]\\
C_{3}(t)&=&{\cal L}^{-1}\left[\frac{1}{d(s)}\left(s^{2}\left(s+\frac{1}{m_{2}}K_{4}(s)\right)+\frac{s}{m_{2}}\left(m_{2}\omega_{2}^{2}-\bigtriangleup_{4}\right)\right)\right]\\
C_{4}(t)&=&-{\cal L}^{-1}\left[\frac{1}{d(s)}\left(\frac{s^{2}}{m_{2}}K_{2}(s)+\frac{s}{m_{2}}\left(g_{R\beta}-\bigtriangleup_{2}\right)\right)\right]\\
D_{1}(t)&=&{\cal L}^{-1}\left[\frac{1}{d(s)}\left(\frac{s}{m_1}\left(\eta
m_1\omega_{1}^{2}-\bigtriangleup_{1}\right)K_{3}(s)-s\left(g_{R\beta}-
\bigtriangleup_{3}\right)\left(s+\frac{1}{m_{1}}K_{1}(s)\right)\right)\right]\\
D_{2}(t)&=&{\cal L}^{-1}\left[\frac{1}{d(s)}\left(\frac{s}{m_{1}}\left(g_{R\beta}-\bigtriangleup_{2}\right)K_{3}(s)-s\left(m_{2}\omega_{2}^{2}-
\bigtriangleup_{4}\right)\left(s+\frac{1}{m_1}K_{1}(s)\right)+\right.\right.\\
&&\left.\left.+\frac{1}{m_1}\left(g_{R\beta}-\bigtriangleup_{2}\right)\left(g_{R\beta}-\bigtriangleup_{3}\right)-\frac{1}{m_{1}}\left(\eta
m_1\omega_{1}^{2}-
\bigtriangleup_{1}\right)\left(m_{2}\omega_{2}^{2}-\bigtriangleup_{4}\right)\right)\right]\\
D_{3}(t)&=&-{\cal L}^{-1}\left[\frac{1}{d(s)}\left(\frac{s^{2}}{m_{1}}K_{3}(s)+\frac{s}{m_{1}}\left(g_{R\beta}-\bigtriangleup_{3}\right)\right)\right]\\
D_{4}(t)&=&{\cal L}^{-1}\left[\frac{1}{d(s)}\left(s^{2}\left(s+\frac{1}{m_{1}}K_{1}(s)\right)+\frac{s}{m_{1}}\left(\eta m_{1}\omega_{1}^{2}-\bigtriangleup_{1}\right)\right)\right]
\end{eqnarray*}
Here ${\cal L}^{-1}$ denotes the inverse Laplace transform, and
$K_{1}(s)$, $K_{2}(s)$, $K_{3}(s)$, and $K_{4}(s)$ are the
Laplace images of the dissipative kernels.

It is convenient to introduce the spectral density $D(\omega)$ of
the heat bath excitations which allows us to replace the sum over
different oscillators $\nu$ by the integral over the frequency:
$\sum\limits_{\nu}^{}\ldots\rightarrow\int\limits_{0}^{\infty}d\omega
D(\omega)\ldots$. This replacement is accompanied by the following
replacements: $\Gamma_{\nu}\rightarrow\Gamma_{\omega}$,
$\omega_{\nu}\rightarrow\omega$ and $n_{\nu}\rightarrow n_{\omega}$.
Let us consider the following spectral functions
\begin{eqnarray}
D(\omega)\frac{|\Gamma(\omega)|^{2}}{\hbar^{2}\omega}=\frac{1}{\pi}\frac{\gamma^2}{\gamma^{2}+\omega^{2}}\nonumber
\end{eqnarray}
where the memory time $\gamma^{-1}$ of the dissipation is inverse to
the phonon bandwidth of the heat bath excitations which are coupled
with the collective oscillator. If we rewrite the sum
$\sum\limits_{\nu}$ as the integral over the bath frequencies with
the density of states, we obtain
\begin{eqnarray}\nonumber
K_{1}(t)=\lambda_1\gamma e^{-\gamma|t|}, \, \, \,
K_{2}(t)=K_{3}(t)=\lambda_1^{1/2}\lambda_{2}^{1/2}\gamma
e^{-\gamma|t|}, \, \, \, K_{4}(t)=\lambda_2\gamma e^{-\gamma|t|}
\end{eqnarray}
and
\begin{eqnarray}\nonumber
\bigtriangleup_{1}=\lambda_{1}\gamma,
\bigtriangleup_{2}=\bigtriangleup_{3}=\lambda_{1}^{1/2}\gamma_{2}^{1/2}\gamma,
\bigtriangleup_{4}=\lambda_{2}\gamma
\end{eqnarray}
We assume that there are no correlations between $F_{1}(t)$ and
$F_{2}(t)$, so that
\begin{eqnarray}
\sum_{\nu}^{}\frac{\alpha_{\nu}g_{\nu}}{\hbar\omega_{\nu}}\equiv0
\end{eqnarray}
The dissipative kernels are  $K_{2}(s)= K_{3}(s)\equiv0$ and $\bigtriangleup_{2}=\bigtriangleup_{3}\equiv0$.
\begin{eqnarray}
K_{1}(s)=\frac{\lambda_{1}\gamma}{(s+\gamma)},
K_{4}(s)=\frac{\lambda_{2}\gamma}{(s+\gamma)}
\end{eqnarray}
So, in this case, the solutions for the collective variables (\ref{inv_lap}) include the following time-dependent coefficients:
\begin{eqnarray*}
A_{1}(t)&=&\frac{1}{m_1m_2}\sum\limits_{i=1}^{6}\xi_{i}\left[m_{1}s_{i}\left(s_{i}+\gamma\right)\left(m_2\left(s_{i}+\gamma\right)\left(s_{i}^{2}+\omega_{2}^{2}\right)-
\lambda_{2}\gamma^{2}\right)+\right.\\
&&+\left.\lambda_{1}\gamma\left(m_{2}\left(s_{i}+\gamma\right)\left(s_{i}^{2}+\omega_{2}^{2}\right)-
\lambda_{2}\gamma^{2}\right)\right]e^{s_{i}t}\\
A_{2}(t)&=&-\frac{g_{R\beta}}{m_1m_2}\sum\limits_{i=1}^{6}\xi_{i}\left[m_{2}s_{i}\left(s_{i}+\gamma\right)^{2}+
\lambda_{2}\gamma\left(s_{i}+\gamma\right)\right]e^{s_{i}t}\\
A_{3}(t)&=&\frac{1}{m_1m_2}\sum\limits_{i=1}^{6}\xi_{i}\left(s_{i}+\gamma\right)\left[m_{2}\left(s_{i}+
\gamma\right)\left(s_{i}^2+\omega_{2}^{2}\right)-\lambda_{2}\gamma^{2}\right]e^{s_{i}t}\\
A_{4}(t)&=&-\frac{g_{R\beta}}{m_1m_2}\sum\limits_{i=1}^{6}\xi_{i}\left(s_{i}+\gamma\right)^{2}e^{s_{i}t}\\
B_{1}(t)&=&-\frac{g_{R\beta}}{m_1m_2}\sum\limits_{i=1}^{6}\xi_{i}\left(s_{i}+\gamma\right)\left[m_{1}s_{i}\left(s_{i}+\gamma\right)+
\lambda_{1}\gamma\right]e^{s_{i}t}\\
B_{2}(t)&=&\frac{1}{m_1m_2}\sum\limits_{i=1}^{6}\xi_{i}\left[m_{2}s_{i}\left(s_{i}+\gamma\right)\left(m_1\left(s_{i}+\gamma\right)\left(s_{i}^{2}+\eta\omega_{1}^{2}\right)-
\lambda_{1}\gamma^{2}\right)+\right.\\
&&+\left.\lambda_{2}\gamma\left(m_{1}\left(s_{i}+\gamma\right)\left(s_{i}^{2}+\eta\omega_{1}^{2}\right)-
\lambda_{1}\gamma^{2}\right)\right]e^{s_{i}t}\\
B_{3}(t)&=&-\frac{g_{R\beta}}{m_1m_2}\sum\limits_{i=1}^{6}\xi_{i}\left(s_{i}+\gamma\right)^{2}e^{s_{i}t}\\
B_{4}(t)&=&\frac{1}{m_1m_2}\sum\limits_{i=1}^{6}\xi_{i}\left(s_{i}+\gamma\right)\left[m_{1}\left(s_{i}+
\gamma\right)\left(s_{i}^2+\eta\omega_{1}^{2}\right)-\lambda_{1}\gamma^{2}\right]e^{s_{i}t}\\
C_{1}(t)&=&\frac{1}{m_2}\sum\limits_{i=1}^{6}\xi_{i}\left(s_{i}+\gamma\right)\left[g_{R\beta}^{2}\left(s_{i}+\gamma\right)-\eta
m_{1}\omega_{1}^{2}\left(m_2\left(s_{i}+\gamma\right)\left(s_{i}^{2}+\omega_{2}^{2}\right)-\lambda_{2}\gamma^{2}\right)+\right.\\
&&+\left.\lambda_{1}\gamma\left(m_2\left(s_{i}+\gamma\right)\left(s_{i}^{2}+\omega_{2}^{2}\right)-\lambda_{2}\gamma^{2}\right)\right]e^{s_{i}t}\\
C_{2}(t)&=&-\frac{g_{R\beta}}{m_2}\sum\limits_{i=1}^{6}\xi_{i}s_{i}\left(s_{i}+\gamma\right)\left(m_2s_{i}(s_{i}+\gamma)+\lambda_{2}\gamma\right)e^{s_{i}t}\\
C_{3}(t)&=&\frac{1}{m_2}\sum\limits_{i=1}^{6}\xi_{i}s_{i}\left(s_{i}+\gamma\right)\left(m_2(s_{i}+\gamma)\left(s_{i}^2+\omega_{2}^2\right)-\lambda_{2}\gamma\right)e^{s_{i}t}\\
C_{4}(t)&=&-\frac{g_{R\beta}}{m_2}\sum\limits_{i=1}^{6}\xi_{i}s_{i}\left(s_{i}+\gamma\right)^{2}e^{s_{i}t}\\
D_{1}(t)&=&-\frac{g_{R\beta}}{m_1}\sum\limits_{i=1}^{6}\xi_{i}s_{i}\left(s_{i}+\gamma\right)\left(m_1s_{i}(s_{i}+\gamma)+\lambda_{1}\gamma\right)e^{s_{i}t}\\
D_{2}(t)&=&\frac{1}{m_1}\sum\limits_{i=1}^{6}\xi_{i}\left(s_{i}+\gamma\right)\left[g_{R\beta}^{2}\left(s_{i}+\gamma\right)-m_{2}\omega_{2}^{2}\left(m_1\left(s_{i}+\gamma\right)
\left(s_{i}^{2}+\eta\omega_{1}^{2}\right)-\lambda_{1}\gamma^{2}\right)+\right.\\
&&+\left.\lambda_{2}\gamma\left(m_1\left(s_{i}+\gamma\right)\left(s_{i}^{2}+\eta\omega_{1}^{2}\right)-\lambda_{1}\gamma^{2}\right)\right]e^{s_{i}t}\\
D_{3}(t)&=&-\frac{g_{R\beta}}{m_1}\sum\limits_{i=1}^{6}\xi_{i}s_{i}\left(s_{i}+\gamma\right)^{2}e^{s_{i}t}\\
D_{4}(t)&=&\frac{1}{m_1}\sum\limits_{i=1}^{6}\xi_{i}s_{i}\left(s_{i}+\gamma\right)\left(m_1(s_{i}+\gamma)\left(s_{i}^2+\eta\omega_{1}^2\right)-\lambda_{1}\gamma\right)e^{s_{i}t}
\end{eqnarray*}
Here, $s_{i}$ are the roots of the following equation:
\begin{eqnarray}
\frac{g_{R\beta}^2\left(s_{i}+\gamma\right)^2}{m_1m_2}-\left(\left(s_{i}+\gamma\right)\left(s_{i}^{2}+\eta\omega_{1}^{2}\right)-
\frac{\lambda_{1}\gamma^2}{m_1}\right)\left(\left(s_{i}+\gamma\right)\left(s_{i}^{2}+\omega_{2}^{2}\right)-\frac{\lambda_{2}\gamma^2}{m_2}\right)=0
\end{eqnarray}
and $\xi_{i}=\left[\prod_{j\neq}\left(s_{i}-s_{j}\right)\right]^{-1}$ with $i,j=1-6$. These roots arise
when we apply the residue theorem to perform integration in
the inverse Laplace transformation.

\subsection{Fluctuation-Dissipation Relations}

An important relation between the dissipation in the dynamics of a
system and the fluctuations in a heat bath with which the system
interacts is the fluctuation-dissipation relation. A first example
of its manifestation is the Nyquist noise in an electric circuit.
This relation is of practical interest in the design of noisy
systems. It is also of theoretical interest in statistical physics
because it is a categorical relation which exists between the
stochastic behavior of many microscopic particles and the
deterministic behavior of a macroscopic system. It is therefore also
useful for the description of the interaction of a system with
fields, such as effects related to radiation reaction and vacuum
fluctuations between atoms and fields in quantum optics. The form of
the fluctuation-dissipation relation is usually given under near-equilibrium conditions via linear response theory. We will see in
this paragraph  that this relation has a much wider scope and a
broader implication than has been understood before. In particular
we want to apply  this relations for problems involving dissipation
kinetic energy  the initial stage of heavy ions collisions.

In \cite{Difc}, fluctuation-dissipation relations were obtained for
(\ref{1_5}),  which connect the macroscopic quantity that describes
dissipation and the microscopic characteristic of the internal
subsystem that expresses fluctuation of random forces. Validity of
these relations means that the dissipative kernels in the
non-Markovian dynamic equations of motion are determined correctly.
The quantum fluctuation-dissipation relation of this form was
obtained in \cite{Katia} and the references therein for the simple
cases of the FC and RWA oscillators. Quantum fluctuation-dissipation
relations differ from classical ones and are reduced to them in the
limit of high temperature $T$ (or when $\hbar \to 0$).

In addition to the temperature fluctuations, the quantum
fluctuations are also considered in them. Since equations of motion
(\ref{1_5}) for the collective coordinates and momenta correspond to
the fluctuation-dissipation relations, our formalism is the basis
for describing quantum statistical effects of collective motion.

We obtain the following relationships for the symmetrized
correlation functions ($k=\alpha, g$) of the random forces
$\varphi^{\nu}_{kk'}(t,t')=\langle\langle
F^{\nu}_{k}(t)F^{\nu}_{k'}(t')+
F^{\nu}_{k'}(t')F^{\nu}_{k}(t)\rangle\rangle$:
\begin{eqnarray}\nonumber
\varphi^{\nu}_{kk'}(t,t')=2k_{\nu}k'_{\nu}\left[2n_{\nu} +
1\right]\cos\left(\omega_{\nu}[t-t']\right)
\end{eqnarray}

Using the properties of random forces, we obtain the quantum
fluctuation--dissipation relations
\begin{eqnarray}
\sum\limits_{\nu}^{}\varphi^{\nu}_{\alpha\alpha}(t,t')\frac{\tanh\left[\frac{\hbar\omega_{\nu}}{2T}\right]}{\hbar\omega_{\nu}}=K_{1}(t-t')\nonumber\\
\sum\limits_{\nu}^{}\varphi^{\nu}_{\alpha g}(t,t')\frac{\tanh\left[\frac{\hbar\omega_{\nu}}{2T}\right]}{\hbar\omega_{\nu}}=K_{2}(t-t')\nonumber\\
\sum\limits_{\nu}^{}\varphi^{\nu}_{g \alpha}(t,t')\frac{\tanh\left[\frac{\hbar\omega_{\nu}}{2T}\right]}{\hbar\omega_{\nu}}=K_{3}(t-t')\nonumber\\
\sum\limits_{\nu}^{}\varphi^{\nu}_{gg}(t,t')\frac{\tanh\left[\frac{\hbar\omega_{\nu}}{2T}\right]}{\hbar\omega_{\nu}}=K_{4}(t-t')
\end{eqnarray}

The validity of the fluctuation--dissipation relationships means that
we correctly specified the dissipative kernels in the non-Markovian
equations of motion.

\section{TRANSPORT COEFFICIENTS}

In order to determine the transport coefficients, we use the
solution (\ref{inv_lap}). Averaging them over the whole system and
taking the time derivative, we obtain the following system of
equations for the first moments:
\begin{eqnarray}\nonumber
\nln \dot{R}(t)\ngn &=&\frac{\nln P(t)\ngn}{m_1} \nonumber\\
\nln \dot{\beta}(t)\ngn &=&\frac{\nln
P_{\beta}(t)\ngn}{m_2}\nonumber\\
\nln \dot{P}(t)\ngn &=&-\lambda_{P}\nln P(t)\ngn  + \rho_{R}\nln P_{\beta}(t)\ngn - c_{R}\nln R(t)\ngn +\delta_{R}\nln \beta(t)\ngn \nonumber\\
\nln \dot{P}_{\beta}(t)\ngn &=&-\lambda_{P_{\beta}}\nln P_{\beta}(t)\ngn  + \rho_{\beta}\nln P(t)\ngn -
c_{\beta}\nln \beta(t)\ngn +\delta_{\beta}\nln R(t)\ngn
\label{fir.moment}
\end{eqnarray}
where the time-dependent coefficients $\lambda_P(t), \lambda_{P_{\beta}}(t), \rho_{R}(t), \rho_{\beta}(t), c_{R}(t), c_{\beta}(t), \delta_{R}(t), \delta_{\beta}(t)$. The coefficients $\lambda_{P, P_{\beta}}(t)$ are related to the friction coefficients. The renormalized stiffnesses are $c_{R,\beta}(t)$. Using Eqs.(\ref{inv_lap}), we write Eqs.(\ref{fir.moment}) for the first moments in which the coefficients after simple algebra are
\begin{eqnarray}
\lambda_{P}(t)&= -\left\{\left[B_1(t)\dot{C}_2(t) -
B_2(t)\dot{C}_1(t)\right]\left[A_3(t)D_4(t)-A_4(t)D_3(t)\right]+\right.\nonumber\\
&\left.+\left[B_1(t)\dot{C}_3(t)-B_3(t)\dot{C}_1(t)\right]
\left[A_4(t)D_2(t)-A_2(t)D_4(t)\right]+\right.\nonumber\\
&\left.+ \left[B_1(t)\dot{C}_4(t) -
B_4(t)\dot{C}_1(t)\right]\left[A_2(t)D_3(t)-A_3(t)D_2(t)\right]+\right.\nonumber\\
&\left.+\left[B_2(t)\dot{C}_3(t) -
B_3(t)\dot{C}_2(t)\right]\left[A_1(t)D_4(t)-A_4(t)D_1(t)\right]+
\right.\nonumber\\
&\left.+\left[B_2(t)\dot{C}_4(t) -
B_4(t)\dot{C}_2(t)\right]\left[A_3(t)D_1(t)-A_1(t)D_3(t)\right]+\right.\nonumber\\
&\left.+\left[B_3(t)\dot{C}_4(t) -
B_4(t)\dot{C}_3(t)\right]\left[A_1(t)D_2(t)-A_2(t)D_1(t)\right]
\right\}/I(t)\nonumber\\
\rho_{R}(t)&= \left\{\left[C_1(t)\dot{C}_2(t) -
C_2(t)\dot{C}_1(t)\right]\left[A_3(t)B_4(t)-A_4(t)B_3(t)\right]+\right.\nonumber\\
&\left.+\left[C_1(t)\dot{C}_3(t)-C_3(t)\dot{C}_1(t)\right]
\left[A_4(t)B_2(t)-A_2(t)B_4(t)\right]+\right.\nonumber\\
&\left.+ \left[C_1(t)\dot{C}_4(t) -
C_4(t)\dot{C}_1(t)\right]\left[A_2(t)B_3(t)-A_3(t)B_2(t)\right]+\right.\nonumber\\
&\left.+\left[C_2(t)\dot{C}_3(t) -
C_3(t)\dot{C}_2(t)\right]\left[A_1(t)B_4(t)-A_4(t)B_1(t)\right]+
\right.\nonumber\\
&\left.+\left[C_2(t)\dot{C}_4(t) -
C_4(t)\dot{C}_2(t)\right]\left[A_3(t)B_1(t)-A_1(t)B_3(t)\right]+\right.\nonumber\\
&\left.+\left[C_3(t)\dot{C}_4(t) -
C_4(t)\dot{C}_3(t)\right]\left[A_1(t)B_2(t)-A_2(t)B_1(t)\right]
\right\}/I(t)\nonumber\\
c_{R}(t)&= -\left\{\left[C_1(t)\dot{C}_2(t) -
C_2(t)\dot{C}_1(t)\right]\left[B_3(t)D_4(t)-B_4(t)D_3(t)\right]+\right.\nonumber\\
&\left.+\left[C_1(t)\dot{C}_3(t)-C_3(t)\dot{C}_1(t)\right]
\left[B_4(t)D_2(t)-B_2(t)D_4(t)\right]+\right.\nonumber\\
&\left.+ \left[C_1(t)\dot{C}_4(t) -
C_4(t)\dot{C}_1(t)\right]\left[B_2(t)D_3(t)-B_3(t)D_2(t)\right]+\right.\nonumber\\
&\left.+\left[C_2(t)\dot{C}_3(t) -
C_3(t)\dot{C}_2(t)\right]\left[B_1(t)D_4(t)-B_4(t)D_1(t)\right]+
\right.\nonumber\\
&\left.+\left[C_2(t)\dot{C}_4(t) -
C_4(t)\dot{C}_2(t)\right]\left[B_3(t)D_1(t)-B_1(t)D_3(t)\right]+\right.\nonumber\\
&\left.+\left[C_3(t)\dot{C}_4(t) -
C_4(t)\dot{C}_3(t)\right]\left[B_1(t)D_2(t)-B_2(t)D_1(t)\right]
\right\}/I(t)\nonumber\\
\delta_{R}(t)=& \left\{\left[C_1(t)\dot{C}_2(t) -
C_2(t)\dot{C}_1(t)\right]\left[A_4(t)D_3(t)-A_3(t)D_4(t)\right]+\right.\nonumber\\
&\left.+\left[C_1(t)\dot{C}_3(t)-C_3(t)\dot{C}_1(t)\right]
\left[A_2(t)D_4(t)-A_4(t)D_2(t)\right]+\right.\nonumber\\
&\left.+ \left[C_1(t)\dot{C}_4(t) -
C_4(t)\dot{C}_1(t)\right]\left[A_3(t)D_2(t)-A_2(t)D_3(t)\right]+\right.\nonumber\\
&\left.+\left[C_2(t)\dot{C}_3(t) -
C_3(t)\dot{C}_2(t)\right]\left[A_4(t)D_1(t)-A_1(t)D_4(t)\right]+
\right.\nonumber\\
&\left.+\left[C_2(t)\dot{C}_4(t) -
C_4(t)\dot{C}_2(t)\right]\left[A_1(t)D_3(t)-A_3(t)D_1(t)\right]+\right.\nonumber\\
&\left.+\left[C_3(t)\dot{C}_4(t) -
C_4(t)\dot{C}_3(t)\right]\left[A_2(t)D_1(t)-A_1(t)D_2(t)\right]
\right\}/I(t)\nonumber\\
I(t)&= \left[B_1(t) D_2(t) -
B_2(t)D_1(t)\right]\left[A_4(t)C_3(t)-A_3(t)C_4(t)\right]+\nonumber\\
&+\left[B_1(t) D_3(t)-B_3(t) D_1(t)\right]
\left[A_2(t)C_4(t)-A_4(t)C_2(t)\right]+\nonumber\\
&+ \left[B_1(t)D_4(t) -
B_4(t)D_1(t)\right]\left[A_3(t)C_2(t)-A_2(t)C_3(t)\right]+\nonumber\\
&+\left[B_2(t)D_3(t) -
B_3(t)D_2(t)\right]\left[A_4(t)C_1(t)-A_1(t)C_4(t)\right]+\nonumber\\
&+\left[B_2(t)D_4(t) -
B_4(t)D_2(t)\right]\left[A_1(t)C_3(t)-A_3(t)C_1(t)\right]+\nonumber\\
&+\left[B_3(t)D_4(t) -
B_4(t)D_3(t)\right]\left[A_2(t)C_1(t)-A_1(t)C_2(t)\right]
\end{eqnarray}
Here, the overdot means the time derivative. The expressions
for the coefficients for the other coordinate are obtained from
these expressions using the following replacements: $A_{i}\leftrightarrow B_{i}$ and
$C_{i}\leftrightarrow D_{i}$ $(i = 1,2,3,4)$.

The equations for the second moments (variances),
\begin{eqnarray}
\sigma_{q_1q_j}(t)=\frac{1}{2}\langle q_{i}(t)q_{j}(t)+q_{j}(t)q_{i}(t)\rangle-\langle q_{i}(t)q_{j}(t)\rangle
\end{eqnarray}
where $q_{i}=R, \beta, P$ or $P_{\beta}$ $(i=1-4)$, are
\begin{eqnarray}
\dot{\sigma}_{RR}(t)&=&\frac{2\sigma_{RP}(t)}{m_1}\nonumber\\
\dot{\sigma}_{\beta\beta}(t)&=&\frac{2\sigma_{RP_{\beta}}(t)}{m_2}\nonumber\\
\dot{\sigma}_{R\beta}(t)&=&\frac{\sigma_{\beta P}(t)}{m_1}+\frac{\sigma_{R P_{\beta}}(t)}{m_2}\nonumber\\
\dot{\sigma}_{RP_{\beta}}(t)&=&-\lambda_{P_{\beta}}\sigma_{RP_{\beta}}(t)+\rho_{\beta}\sigma_{RP}(t)-
c_{\beta}\sigma_{R\beta}(t)+\delta_{\beta}\sigma_{RR}(t)+\frac{\sigma_{PP_{\beta}(t)}}{m_1}+2D_{RP_{\beta}}(t)\nonumber\\
\dot{\sigma}_{RP}(t)&=&-\lambda_{P}\sigma_{RP}(t)+\rho_{R}\sigma_{RP_{\beta}}(t)-
c_{R}\sigma_{RR}(t)+\delta_{R}\sigma_{R\beta}(t)+\frac{\sigma_{PP}(t)}{m_1}+2D_{RP}(t)\nonumber\\
\dot{\sigma}_{\beta P}(t)&=&-\lambda_{P}\sigma_{\beta
P}(t)+\rho_{R}\sigma_{\beta P_{\beta}}(t)-
c_{R}\sigma_{R\beta}(t)+\delta_{R}\sigma_{\beta\beta}(t)+\frac{\sigma_{PP_{\beta}}(t)}{m_2}+2D_{\beta
P}(t)\nonumber\\
\dot{\sigma}_{\beta
P_{\beta}}(t)&=&-\lambda_{P_{\beta}}\sigma_{\beta
P_{\beta}}(t)+\rho_{\beta}\sigma_{\beta P}(t)-
c_{\beta}\sigma_{\beta\beta}(t)+\delta_{\beta}\sigma_{R\beta}(t)+\frac{\sigma_{P_{\beta}P_{\beta}(t)}}{m_2}+2D_{\beta P_{\beta}}(t)\nonumber\\
\dot{\sigma}_{PP_{\beta}}(t)&=&-\left(\lambda_P+\lambda_{P_{\beta}}\right)\sigma_{PP_{\beta}}(t)+\rho_{R}\sigma_{P_{\beta}P_{\beta}}(t)+
\rho_{\beta}\sigma_{PP}(t)-c_R\sigma_{RP_{\beta}}(t)-c_{\beta}\sigma_{\beta P}(t)+\nonumber\\
&+&\delta_{R}\sigma_{\beta P_{\beta}}(t)+\delta_{\beta}\sigma_{RP}(t)+2D_{PP_{\beta}}(t)\nonumber\\
\dot{\sigma}_{P_{\beta}P_{\beta}}(t)&=&-2\lambda_{P_{\beta}}\sigma_{P_{\beta}P_{\beta}}(t)+2\rho_{\beta}\sigma_{PP_{\beta}}(t)-2c_{\beta}\sigma_{\beta P_{\beta}}(t)+2\delta_{\beta}\sigma_{RP_{\beta}}(t)+2D_{P_{\beta}P_{\beta}}(t)\nonumber\\
\dot{\sigma}_{PP}(t)&=&-2\lambda_{P}\sigma_{PP}(t)+2\rho_{R}\sigma_{PP_{\beta}}(t)-2c_{R}\sigma_{RP}(t)+2\delta_{R}\sigma_{\beta
P}(t)+2D_{PP}(t)
\end{eqnarray}
So we have obtained the Markovian-type (local in time)
equations for the first and second moments, but with the
transport coefficients depending explicitly on time.The time-dependent
diffusion coefficients $D_{q_{i}q_{j}}(t)$  are determined as
\begin{eqnarray}\nonumber
D_{RR}(t)&=&-\frac{J_{RP}(t)}{m_1}+\frac{1}{2}\dot{J}_{RR}(t)\nonumber\\
D_{\beta\beta}(t)&=&-\frac{J_{\beta P_{\beta}}(t)}{m_2}+\frac{1}{2}\dot{J}_{\beta\beta}(t)\nonumber\\
D_{R\beta}(t)&=&-\frac{1}{2}\left[\frac{J_{\beta
P}}(t){m_1}+\frac{J_{RP_{\beta}}(t)}{m_2}-\dot{J}_{R\beta}(t)\right]\nonumber\\
D_{RP_{\beta}}(t)&=&-\frac{1}{2}\left[-\lambda_{P_{\beta}}J_{RP_{\beta}}(t)+\rho_{\beta}J_{RP}(t)-
c_{\beta}J_{R\beta}(t)+\delta_{\beta}J_{RR}(t)+\frac{J_{PP_{\beta}}(t)}{m_1}-\dot{J}_{RP_{\beta}}(t)\right]\nonumber\\
D_{RP}(t)&=&-\frac{1}{2}\left[-\lambda_{P}J_{RP}(t)+\rho_{R}J_{RP_{\beta}}(t)-
c_{R}J_{RR}(t)+\delta_{R}J_{R\beta}(t)+\frac{J_{PP}(t)}{m_1}-\dot{J}_{RP}(t)\right]\nonumber\\
D_{\beta P}(t)&=&-\frac{1}{2}\left[-\lambda_{P}J_{\beta
P}(t)+\rho_{R}J_{\beta P_{\beta}}(t)-
c_{R}J_{RR}(t)+\delta_{R}J_{\beta\beta}(t)+\frac{J_{PP_{\beta}}(t)}{m_2}-\dot{J}_{\beta P}(t)\right]\nonumber\\
D_{\beta
P_{\beta}}(t)&=&-\frac{1}{2}\left[-\lambda_{P_{\beta}}J_{\beta
P_{\beta}}(t)+\rho_{\beta}J_{\beta P}(t)-
c_{\beta}J_{\beta\beta}(t)+\delta_{\beta}J_{R\beta}(t)+\frac{J_{P_{\beta}P_{\beta}}(t)}{m_2}-\dot{J}_{\beta P_{\beta}}(t)\right]\nonumber\\
D_{PP_{\beta}}(t)&=&-\frac{1}{2}\left[-\left(\lambda_{P}+\lambda_{P_{\beta}}\right)J_{PP_{\beta}}(t)+\rho_{R}J_{P_{\beta}P_{\beta}}(t)+\rho_{\beta}J_{PP}(t)-
c_{R}J_{RP_{\beta}}(t) -c_{\beta}J_{\beta P}(t)+\right.\nonumber\\
&+&\left.\delta_{R}J_{\beta P_{\beta}}(t)+\delta_{\beta}J_{RP}(t)-\dot{J}_{PP_{\beta}}(t)\right]\nonumber\\
D_{P_{\beta}P_{\beta}}(t)&=&\lambda_{P_{\beta}}J_{P_{\beta}P_{\beta}}(t)-\rho_{\beta}J_{PP_{\beta}}(t)+
c_{\beta}J_{\beta P_{\beta}}(t)-\delta_{\beta}J_{RP_{\beta}}(t)+\frac{1}{2}\dot{J}_{P_{\beta}P_{\beta}}(t)\nonumber\\
D_{PP}(t)&=&\lambda_{P}J_{PP}(t)-\rho_{R}J_{PP_{\beta}}(t)+c_{R}J_{RP}(t)-\delta_{R}J_{\beta
P}(t)+\frac{1}{2}\dot{J}_{PP}(t)
\label{D123}
\end{eqnarray}
Here, $\dot{J}_{q_{i}q_{j}}(t)=dJ_{q_{i}q_{j}}(t)/dt$. In our
treatment $D_{RR}=0$, $D_{\beta\beta}=0$, and $D_{R\beta}=0$ because
there are no random forces for the $R$ and $\beta$ coordinates in
Eqs. (\ref{1_5}). In Eqs. (\ref{D123}) we use the following
notation:
\begin{eqnarray}\nonumber
J_{RR}(t)&=&\langle\langle
I_{R}(t)I_{R}(t)+I_{R}'(t)I_{R}'(t)\rangle\rangle,\nonumber\\
J_{\beta\beta}(t)&=&\langle\langle
I_{\beta}(t)I_{\beta}(t)+I_{\beta}'(t)I_{\beta}'(t)\rangle\rangle,\nonumber\\
J_{PP}(t)&=&\langle\langle
I_{P}(t)I_{P}(t)+I_{P}'(t)I_{P}'(t)\rangle\rangle,\nonumber\\
J_{P_{\beta}P_{\beta}}(t)&=&\langle\langle
I_{P_{\beta}}(t)I_{P_{\beta}}(t)+I_{P_{\beta}}'(t)I_{P_{\beta}}'(t)\rangle\rangle,\nonumber\\
J_{PP_{\beta}}(t)&=&\langle\langle
I_{P}(t)I_{P_{\beta}}(t)+I_{P}'(t)I_{P_{\beta}}'(t)\rangle\rangle,\nonumber\\
J_{R\beta}(t)&=&\langle\langle
I_{R}(t)I_{\beta}(t)+I_{R}'(t)I_{\beta}'(t)\rangle\rangle,\nonumber\\
J_{RP}(t)&=&\langle\langle
I_{R}(t)I_{P}(t)+I_{R}'(t)I_{P}'(t)\rangle\rangle,\nonumber\\
J_{\beta P_{\beta}}(t)&=&\langle\langle
I_{\beta}(t)I_{P_{\beta}}(t)+I_{\beta}'(t)I_{P_{\beta}}'(t)\rangle\rangle,\nonumber\\
J_{R P_{\beta}}(t)&=&\langle\langle
I_{R}(t)I_{P_{\beta}}(t)+I_{R}'(t)I_{P_{\beta}}'(t)\rangle\rangle,\nonumber\\
J_{\beta P}(t)&=&\langle\langle
I_{\beta}(t)I_{P}(t)+I_{\beta}'(t)I_{P}'(t)\rangle\rangle.
\end{eqnarray}

Thus, we obtain equations for the first and second moments with the
transport coefficients explicitly depending on time, collective
coordinate, and momentum. It is the time dependence of these
coefficients that results from the non-Markovian nature of the
system.

\section{CONCLUSIONS}
A system of nonlinear Langevin equations is derived within the
microscopic approach in the limit of the general coupling between
the collective and internal  subsystems. These equations of motion
for the collective subsystem satisfy the quantum
fluctuation-dissipation relations. A new method for obtaining
explicitly time-dependent transport coefficients is developed on the
basis of the non-Markovian Langevin equations.  The analytical
formulas obtained in this work can be used for describing the
fluctuation-dissipation dynamics of nuclear processes.



\begin{thebibliography}{99}

%
\bibitem{Difc}           {\it Kanokov Z., Palchikov Yu. V., Adamian G. G., Antonenko N. V., Scheid W.} Non-Markovian dynamics of quantum systems.
                         I. Formalism and transport coefficients // Phys. Rev. E. 2005. V. 71. P. 016121;
                         {\it Palchikov Yu. V., Kanokov Z., Adamian G. G., Antonenko N. V., Scheid W.} Non-Markovian dynamics of quantum systems.
                         II. Decay rate, capture, and pure states // Phys. Rev. E. 2005. V. 71. P. 016122.
\bibitem{Kalan}          {\it Kalandarov Sh. A., Kanokov Z.,Adamian G. G., Antonenko N. V.} Influence of external magnetic field on dynamics of open
                         quantum systems // Phys. Rev. E. 2007. V. 75. P. 031115.
\bibitem{MyAllDM}       {\it Sargsyan V. V., Palchikov Yu. V., Kanokov Z., Adamian G. G., Antonenko N. V.} Coordinate-dependent diffusion coefficients: Decay rate in open quantum
                        systems // Phys. Rev. A. 2007. V. 75. P. 062115;
                        {\it Sargsyan V. V., Palchikov Yu. V., Kanokov Z., Adamian G. G., Antonenko N. V.} Fission rate and transient time with a quantum master equation // Phys.
                         Rev. C. 2007. V. 76. P. 064604.
\bibitem{MyAllQDA}       {\it Sargsyan V. V., Adamian G. G., Antonenko N. V., Scheid W.} Peculiarities of sub-barrier fusion with quantum diffusion approach // Eur. Phys. J.
                         A. 2010. V. 45. P. 125;
                         {\it Sargsyan V. V., Adamian G. G., Antonenko N. V., Scheid W., Zhang, H. Q.} Sub-barrier capture with quantum diffusion approach: Actinide-based reactions //
                         Eur. Phys. J. A. 2011. V. 47. P. 38.
\bibitem{MyAllNT}        {\it Sargsyan V. V., Adamian G. G., Antonenko N. V., Scheid W., Zhang, H. Q.} Effects of nuclear deformation and neutron transfer in capture processes,
                         and fusion hindrance at deep sub-barrier energies // Phys. Rev. C. 2011. V. 84. P. 064614;
                         {\it Sargsyan V. V., Adamian G. G., Antonenko N. V., Scheid W., Zhang, H. Q.} Role of neutron transfer in capture processes at sub-barrier energies // Phys.
                         Rev. C. 2012. V. 85. P. 024616.
\bibitem{obzor}          {\it Adamian G. G., Nasirov A. K., Antonenko N. V., Jolos R. V.} The influence of the shell effects on dynamics of deep-inelastic
                         colisions of heavy ions // Phys. Part. Nucl. 1994. V. 25. P. 583.
\bibitem{Leggett}        {\it Caldeira A. O.,  Leggett A. J.} Path integral approach to quantum Brownian motion // Physica A. 1983. V. 121. P. 587;
                         Quantum tunnelling in a dissipative system // Ann. Phys. 1983. V.149. P.374;
                         Influence of dissipation on quantum tunneling in macroscopic systems // Phys. Rev. Lett. 1981. V. 46. P. 211;
                         Comment on "Probabilities for Quantum Tunneling through a Barrier with Linear Passive Dissipation" // Phys. Rev. Lett. 1982. V. 48. P. 1571.
\bibitem{Katia}          {\it K. Lindenberg, B. West}. Phys. Rev. A. 1984. V.30, P.568.
\bibitem{Weiss}          {\it Weiss U.} Quantum Dissipative Systems. Singapore: Wold Scientific, 1999.
\bibitem{Lindblad}       {\it G. Lindblad}. Commun. Math. Phys. 1976. V.48, P.119;
                         Rep. Math. Phys. 1976. V.10, P.393.
\bibitem{Dorofeyev}      {\it Illarion Dorofeyev}. ArXiv. 1207.3881.
\bibitem{MyYAF1}         {\it V. V. Sargsyan et al.}. Yad. Fiz. 2005. V.68, P.2071.
\bibitem{MyTMF1}         {\it G. G. Adamyan et al.}. Teor. Mat. Fiz. 2005. V.145, P.87.
\bibitem{Kampen}         {\it van Kampen N. G.} Stochastic Processes in Physics and Chemistry. Amsterdam: North-Holland, 1981.
\bibitem{Gardiner}       {\it Gardiner C. W.} Quantum Noise. Berlin: Springer, 1991.



\end{thebibliography}
\end{document}